\documentclass[12pt]{article}
\usepackage[
  top=1in,
  bottom=1in,
  left=1in,
  right=1in
]{geometry}
\usepackage{amsmath}
\usepackage{graphicx}
\usepackage{enumerate}
\usepackage{natbib}
\usepackage{url} 
\usepackage{amsfonts,amstext,amssymb}
\usepackage{bbm}
\usepackage{mathtools}
\usepackage{algorithm}
\usepackage{algpseudocode}
\usepackage{fancyvrb}
\usepackage{tikz}
\usepackage{subfigure}
\usepackage{multirow}
\usepackage{tabularx}
\usepackage{capt-of}
\usepackage{mathdots} 
\usepackage{multirow}
\usepackage{bigstrut}
\usepackage{booktabs}
\usepackage{caption}
\usepackage{ulem}
\usepackage{mathtools}
\usepackage{hyperref}
\usepackage{authblk}
\usepackage{setspace} 
\DeclarePairedDelimiter\ceil{\lceil}{\rceil}

\def\bSig\mathbf{\Sigma}

\newcommand{\abs}[1]{\lvert #1 \rvert}

\newcommand{\keywords}[1]{%
  \vspace{1em}
  \noindent\textbf{Keywords:} #1
}

\renewcommand{\footnoterule}{%
  \kern -3pt
  \hrule width \textwidth height 0.4pt
  \kern 2pt
}

\makeatletter
\renewcommand{\@maketitle}{
  \newpage
  \null
  \vskip 1em%
  \begin{center}%
    {\LARGE \@title \par}%
    \vskip 1em%
    {\large
      \lineskip .5em%
      \begin{tabular}[t]{c}%
        \@author
      \end{tabular}\par}%
    \vskip 0.5em%
  \end{center}%
  \par
  \vskip 0.5em
}
\makeatother

\title{{Testing-driven Variable Selection in Bayesian Modal Regression}\thanks{This manuscript is a preprint version of the article currently under review. It is distributed under the arXiv.org perpetual, non-exclusive license.}}

\author{
  Jiasong Duan$^{1,*}$, Hongmei Zhang$^{2,**}$, Xianzheng Huang$^{1,***}$\\
  {\small $^{1}$Department of Statistics, University of South Carolina, Columbia, South Carolina, U.S.A.}\\
  {\small $^{2}$Division of Epidemiology, Biostatistics, and Environmental Health Sciences, \\
 School of Public Health, University of Memphis, Memphis, Tennessee, U.S.A.}\\
  {\small $^{*}$\textit{email:} jiasong@email.sc.edu}\\
  {\small $^{**}$\textit{email:} hongmei.zhang@memphis.edu}\\
  {\small $^{***}$\textit{email:} huang@stat.sc.edu}
}
\date{} 

\begin{document}

\maketitle

\begin{abstract}
\noindent 
We propose a Bayesian variable selection method in the framework of modal regression for heavy-tailed responses. An efficient expectation-maximization algorithm is employed to expedite parameter estimation. A test statistic is constructed to exploit the shape of the model error distribution to effectively separate informative covariates from unimportant ones. Through simulations, we demonstrate and evaluate the efficacy of the proposed method in identifying important covariates in the presence of non-Gaussian model errors. Finally, we apply the proposed method to analyze two datasets arising in genetic and epigenetic studies.  
\end{abstract}


\section{Introduction}\label{sec:intro}
Variable selection has been the subject of extensive research in the statistics and machine learning communities for decades. The goal is to select informative predictors as covariates in a regression model for the response of interest, such as identifying genes as markers for a particular health condition, discovering significant contributors to greenhouse gas concentrations, or finding contributing factors that impact the price of insurance premiums. From a statistical standpoint, screening out unimportant covariates in a regression model helps avoid overfitting and enhances interpretability. 

Stepwise selection methods \citep[][Section 3.3]{hastie2017elements}, including forward selection and backward selection, are well-received methods for variable selection when the number of candidate covariates is not large. Over the years, the demand for analyzing high-dimensional data has driven advancements in variable selection methods that can handle a large number of candidate covariates. These include frequentist shrinkage methods based on penalized objective functions or score functions \citep{tibshirani1996regression,fan2001variable,zou2005regularization,zou2006adaptive, huang2013variable}, or using penalized estimating equations \citep{ma2010variable}. Methods for variable selections under the Bayesian framework typically utilize prior distributions for regression coefficients along with indicator variables for the inclusion or exclusion of covariates \citep{george1993variable, rovckova2014emvs, zhang2016two, rovckova2018spike}. In particular, \citet{rovckova2014emvs} 
used a spike-and-slab prior on covariate effects, and applied the expectation-maximization (EM) algorithm to speed up parameter estimation, following which negligible regression coefficients are filtered out using a specific threshold. The authors later proposed a method called spike-and-slab LASSO \citep[SSL,][]{rovckova2018spike} that can obtain exactly zero estimates for negligible coefficients to achieve simultaneous coefficient estimation and variable selection. The efficacy of SSL heavily depends on two hyperparameters in the spike-and-slab prior \citep{bai2021spike}, as is the case for many other Bayesian variable selection methods with different formulations of priors. This is one complication of Bayesian variable selection that our study addresses, besides a practical issue regarding the response data. 

Most existing variable selection methods were developed for mean regression models that assume Gaussian response data conditional on covariates. Non-Gaussian response data and data subject to severe outliers are ubiquitous in many applications. For instance, more than half of gene expression
data are not normally distributed \citep{de2020shape}; patients' wait times for elective surgeries can exhibit heavy tails due to excessive waits in a few complex cases \citep{kraus2024equal}; and extreme outliers in the size of wildfire are present because of the few reaching catastrophic sizes under rare but extreme weather conditions \citep{schoenberg2003distribution}. Existing variable selection methods suitable for Gaussian responses can lead to suboptimal performance when confronted with non-Gaussian response data. Such concern can be eased through variable selection based on quantile regression models \citep{li20081,Zou2008Composite,wu2009variable}. Although quantile regression is more robust to outliers than mean regression, the interpretability of selection outcomes is somewhat compromised because they are quantile-specific. An alternative to relieve the concern of normality violation is to utilize nonparametric approaches  \citep{breiman2001random,Chipman2010BART,huang2010variable}. However, the price they pay is statistical efficiency and interpretability of the selected model. 

In this study, we propose a Bayesian variable selection method under a parametric framework where regression toward the mode of the response is considered. Embedding the covariate effects in the conditional mode of the response guarantees robustness to outliers in inferences for covariate effects. Furthermore, we assume for the error distribution of the regression model a flexible unimodal distribution that can be light-tailed or heavy-tailed, symmetric or asymmetric. This regression framework is presented in Section~\ref{sec:The regression model}. Section~\ref{sec:Parameter estimation} provides a detailed account of parameter estimation under the Bayesian modal regression framework, including prior elicitation and the EM algorithm for a more numerically efficient update of estimates than a conventional Markov chain Monte Carlo (MCMC) algorithm. Following parameters estimation, we propose in Section~\ref{sec:variable selection} a test statistic motivated by the shape of the unimodal error distribution to screen out negligible regression coefficients. Section~\ref{sec:Simulations} presents simulation studies where we compare the proposed strategy with two highly relevant competing methods. These methods are then applied to two datasets in the medical field in Section~\ref{sec:Real data examples}. Section~\ref{sec:disc} summarizes our contribution and outlines future research directions. 

\section{The modal regression model}\label{sec:The regression model}
Denote by $\boldsymbol{X}=(\boldsymbol{x}_1,  \ldots, \boldsymbol{x}_n)^\top$ an $n \times p$ matrix consisting of measurements of $p$ covariates associated with $n$ subjects, and by  $\boldsymbol{y}=(y_1, \ldots, y_n)^\top$ the response data of the $n$ subjects. Given $\boldsymbol{X}$, we assume a model for $\boldsymbol{y}$ specified by the following linear model,
\begin{equation}\label{eq:Linear Regression}
 \boldsymbol y= \beta_0 \boldsymbol{1}_n+ \boldsymbol X \boldsymbol \beta + \boldsymbol \varepsilon,
\end{equation} 
where $\boldsymbol{1}_n$ is the $n\times 1$ vector of ones, $\beta_0$ is the intercept, $\boldsymbol \beta =(\beta_1, \ldots, \beta_p)^\top$ is the vector of covariate effects, and $\boldsymbol{\varepsilon}=(\varepsilon_1, \ldots, \varepsilon_n)^\top$ is a vector of independent and identically distributed random errors from a mode-zero distribution. By construction, \eqref{eq:Linear Regression} reveals a regression function defined by the mode of the response as a linear function of the covariates. Modal regression has been much less explored than mean regression and quantile regression, although there has been some revival of interest in modal regression recently thanks to its robustness to outliers and efficiency in prediction \citep{chacon2020modal}. Regardless, compared to variable selection in mean regression and quantile regression, variable selection in modal regression remains underdeveloped. The limited existing works on this topic are mostly in the nonparametric or semiparametric framework  \citep{zhang2013robust, wang2017regularized, guo2019robust, wang2019modal, xia2022variable} owing to the lack of familiar distribution families that highlight the mode as a location parameter. This signifies the contribution of our proposed regression model with a flexible error distribution, of which the mode is zero. 

\begin{figure}[htbp]
    \centering
    \subfigure[Densities of $\varepsilon_i$ with different $\nu$ values]{
        \includegraphics[width=0.45\textwidth]{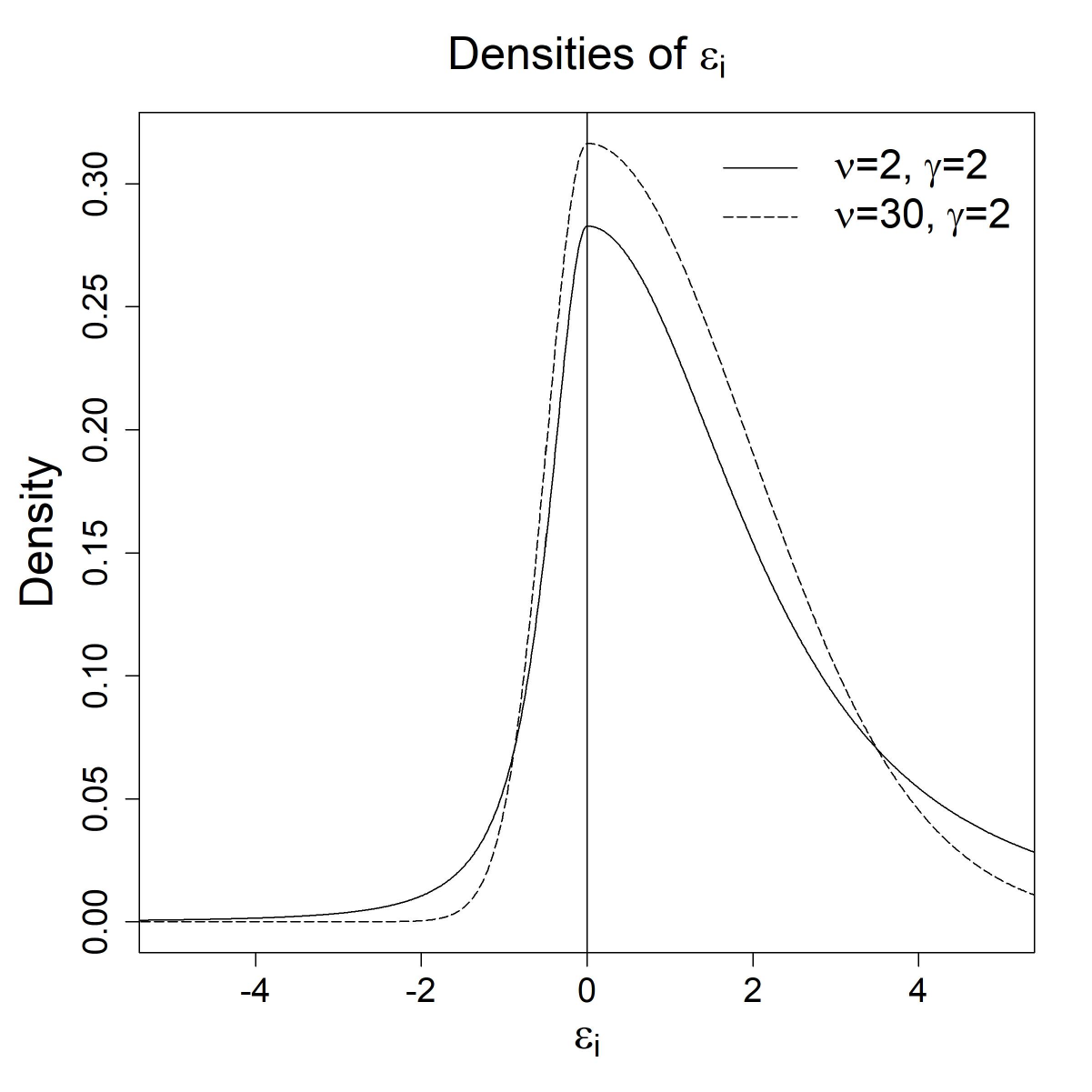}
        \label{fig:figure1a}
    }
    \hfill
    \subfigure[Densities of $\varepsilon_i$ with different $\gamma$ values]{
        \includegraphics[width=0.45\textwidth]{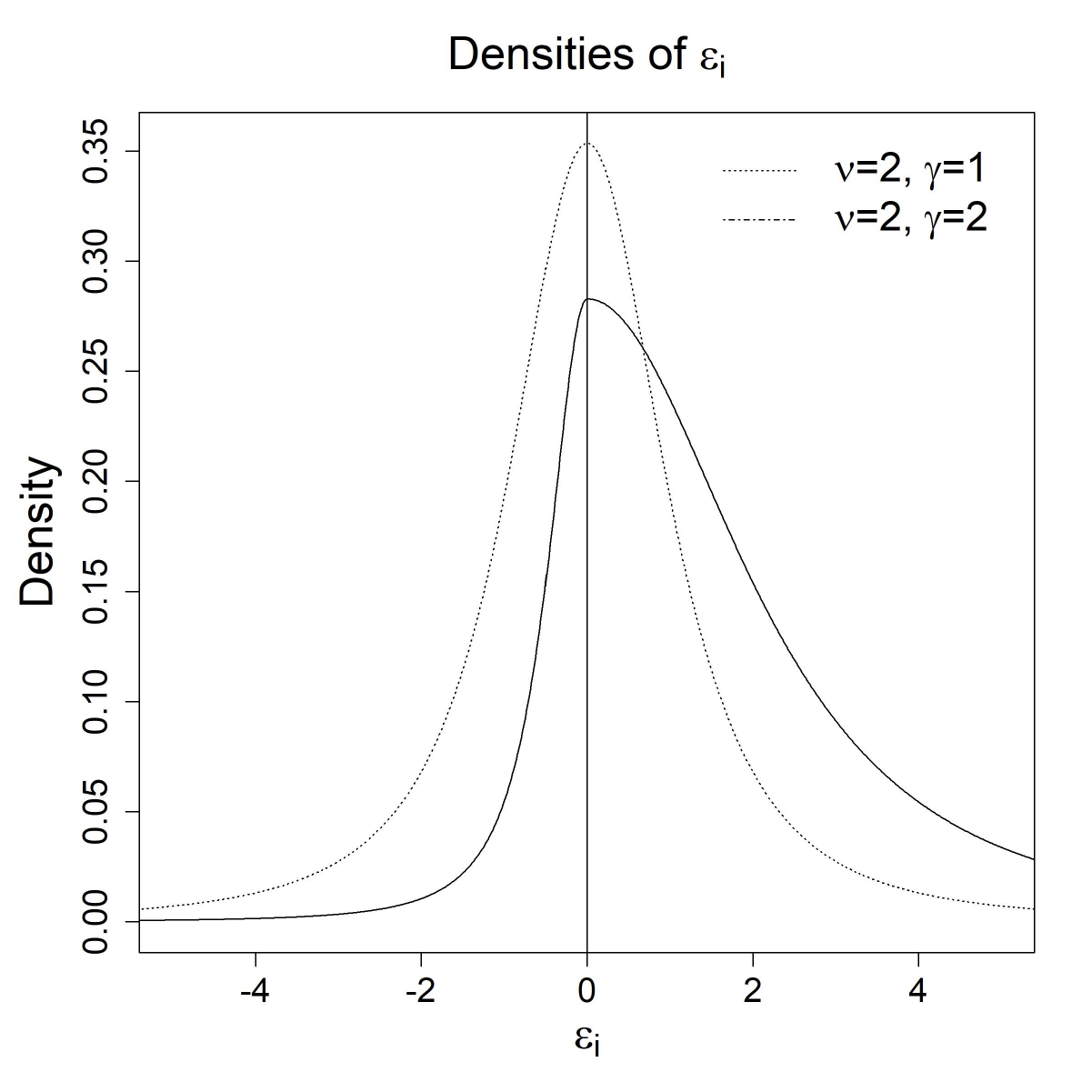}
        \label{fig:figure1b}
    }
    \caption{Probability density functions of $\varepsilon_i$ with parameters $\nu$ and $\gamma$ set at different values}
    \label{fig:densities with different parameters}
\end{figure}

To model a response potentially subject to extreme outliers and asymmetry, we exploit the unimodal two-component mixture distribution introduced by   \cite{fernandez1998bayesian} to characterize the model error in \eqref{eq:Linear Regression}, with the probability density function given by
\begin{equation}
\label{eq:error distribution}
p(\varepsilon_i|\nu, \gamma) = \frac{2}{\gamma+\gamma^{-1}}\left\{ f_{\nu}\left(\frac{\varepsilon_i}{\gamma}\right)\mathbbm{1}_{[0, +\infty)}(\varepsilon_i)+ f_{\nu}(\gamma\varepsilon_i)\mathbbm{1}_{(-\infty, 0)}(\varepsilon_i) \right\},
\end{equation}
where $\gamma>0$ is a shape parameter, $\mathbbm{1}_A(\varepsilon_i)$ is the indicator function returning one if $\varepsilon_i \in A$, and $f_\nu(\cdot)$ is the probability density function of a unimodal distribution that is symmetric around zero and involves parameter(s) $\nu$. For a concrete exposition of the proposed method, we set $f_\nu(\cdot)$ as the density of a Student's $t$ distribution with $\nu$ degrees of freedom, referred to as $t_\nu$. With this choice, $\nu$ determines the heaviness of the tail of $f_\nu(\cdot)$, and thereby, that of $p(\varepsilon_i|\nu, \gamma)$.  
The shape parameter $\gamma$ controls the allocation of probability masses on either side of mode zero of the error distribution. Setting $\gamma=1$ yields an error distribution symmetric about zero; a larger (smaller) value of $\gamma$ gives rise to a distribution skewing to the right (left) more. Because \eqref{eq:error distribution} is essentially a mixture of the right half of $t_\nu$ and the left half $t_\nu$ evaluated at differently scaled (unless $\gamma=1$) error, we call the distribution specified by \eqref{eq:error distribution} a \underline{mix}ture of \underline{ha}lf \underline{$t$} distribution, or MixHat for short, and say that $\varepsilon_i\sim \text{MixHat}(\nu, \gamma)$.   
Figure \ref{fig:densities with different parameters} presents the density of  $ \text{MixHat}(\nu, \gamma)$ with different values set for $(\nu, \gamma)$. These 
examples demonstrate the flexibility of the error distribution, which in turn leads to a flexible model for the response according to \eqref{eq:Linear Regression} that can capture heavy-tailed or thin-tailed, skewed or symmetric response data. 

Unknown parameters in the modal regression model specified by \eqref{eq:Linear Regression} and \eqref{eq:error distribution} include $\beta_0, \boldsymbol{\beta}, \nu$ and $\gamma$. We estimate these parameters in the framework of Bayesian inferences. Variable selection is facilitated by encouraging sparsity in $\boldsymbol \beta$.

\section{Parameter estimation under Bayesian framework}\label{sec:Parameter estimation}
\subsection{Prior and posterior distributions}\label{subsec:prior distributions}
In this study, we adopt the spike-and-slab prior for $\boldsymbol \beta$, formulated as follows \citep{george1993variable,rovckova2018SSL}, 
\begin{equation}
\begin{aligned}
\label{eq:BetaPrior}
\pi\left(\boldsymbol{\beta} \mid  \boldsymbol{\lambda}\right) & =\prod_{j=1}^p \Bigl\{ (1-\lambda_j)\psi(\beta_j \mid t_0) +\lambda_j\psi(\beta_j \mid t_1)\Bigl\}, \\
\pi\left(\boldsymbol{\lambda} \mid  \theta \right) & = \prod_{j=1}^p \theta^{ \lambda_j}(1-\theta)^{1- \lambda_j}, \\
\pi\left(\theta\right) & \propto \theta^{a-1}(1-\theta)^{b-1}, 
\end{aligned}
\end{equation}
where $\psi(s \mid t)=(t/2)\exp(-t \abs{s})$ is the density of a Laplace distribution with mean zero and rate parameter $t$, $\boldsymbol{\lambda}=(\lambda_1, \ldots, \lambda_p)^\top$ contains $p$ binary latent variables, with $\lambda_j=1$ indicating the inclusion of the $j$-th covariate and $\lambda_j=0$ for exclusion, for $j=1, \ldots, p$. The hyper-prior distribution for $\boldsymbol \lambda$ originates from independent Bernoulli priors for $\lambda_j$
's, with a success probability of $\theta$, for which the hyper-prior distribution is a beta distribution with positive hyperparameters $a$ and $b$, i.e., $\text{beta}(a, b)$. We set $(a, b)=(1, p)$ following the recommendation in \cite{castillo2012needles}, resulting in the prior for $\theta$ as $\text{beta}(1, p)$ that has a smaller mean and variance as $p$ increases.
Consequently, as the number of candidate covariates becomes larger, it suggests a priori a lower probability of each covariate being selected, coupled with a reduced uncertainty of the selection, which makes this prior especially suitable when sparsity is expected. The rate parameters $t_0$ and $t_1$ in \eqref{eq:BetaPrior} are crucial in Bayesian variable selection \citep{george1997approaches,rovckova2014emvs,rovckova2018spike}. A routinely implemented strategy is to set $t_1$ much smaller than $t_0$ in order to achieve a much higher variance (due to a small $t_1$) for the ``slab'' portion and a lower variance (due to a large $t_0$) for the ``spike" part of the prior for $\boldsymbol{\beta}$.

For the remaining parameters not directly related to variable selection, $(\beta_0, \nu, \gamma)$, we adopt well-established priors. For $\beta_0$, a vague Gaussian prior with a mean of zero and a variance of $10^6$, $\mathcal{N}(0, 10^6)$, is used. For $\nu$, the prior is the log-normal distribution with the logarithm of location and scale parameters both equal to 1 as recommended by \cite{lee2022use}. This prior assigns most probability mass to the range of $(0,25)$ that yields a wide variety of $t$ distributions, including $t_\nu$ with much heavier tails than Gaussian, as well as tails light enough for $t_\nu$ to resemble $\mathcal{N}(0, 1)$.
For $\gamma$, we use a gamma distribution with the shape and rate parameter both equal to $10^{-4}$ for a relatively non-informative prior. 

With the introduction of priors and hyper-priors, the collection of unknown parameters is expanded to $\Theta=(\beta_0, \boldsymbol \beta^\top,  \nu, \gamma, \theta, \boldsymbol{\lambda}^\top)^\top$. The joint posterior distribution of $\Theta$, up to a normalizing constant, is as follows, 
\begin{equation}
\label{eq:JointPosterior}
\begin{aligned}
p(\Theta \mid \boldsymbol{y}) 
 \propto &\ f(\boldsymbol{y} \mid \beta_0, \boldsymbol{\beta},  \nu, \gamma)\pi(\beta_0)\pi(\boldsymbol{\beta} \mid \boldsymbol{\lambda}) \pi\left(\boldsymbol{\lambda} \mid  \theta \right) \pi(\theta) 
  \pi(\nu) \pi(\gamma)  \\
 \propto &\ \prod_{i=1}^n \left\{ \left(1+\frac{\varepsilon_i^2}{\gamma^2 \nu} \right)^{-(\nu+1)/2} \mathbbm{1}_{[0,+\infty)}(\varepsilon_i)+ \left(1+\frac{\gamma^2\varepsilon_i^2}{\nu} \right)^{-(\nu+1)/2} \mathbbm{1}_{(-\infty,0)}(\varepsilon_i)\right\}\times\\
&\ e^{-\beta_0^2/(2\times 10^6)} \prod_{j=1}^p \left\{ \left(\frac{t_0}{2}e^{-t_0|\beta_j|}\right)^{1-\lambda_j} \left(\frac{t_1}{2}e^{-t_1|\beta_j|}\right)^{\lambda_j} \right\} \displaystyle{\frac{\theta^{\sum_{j=1}^p \lambda_j + a - 1}}{(1 - \theta)^{\sum_{j=1}^p \lambda_j - b + 1-p}}}\times  \\
&\ \frac{e^{-(\log \nu-1)^2/2}}{\nu^{n/2+1}}
\left\{ \frac{\Gamma\bigl((\nu + 1)/2)}{\Gamma(\nu/2)} \right\}^n \frac{\gamma^{n + c - 1}}{(\gamma^2 + 1)^n} e^{-d\gamma},
\end{aligned}
\end{equation}
where $\varepsilon_i=y_i-\beta_0-\boldsymbol{x}_i^\top \boldsymbol{\beta}$, for $i=1, \ldots, n$, $a=1$, $b=p$, and $c=d=10^{-4}$. 

\subsection{The expectation-maximization algorithm}\label{subsec:E-M algorithm}

To improve efficiency over the classical MCMC simulations, we opt for the expectation-maximization (EM) algorithm to obtain maximum a posteriori estimates, which, in general, converges faster than MCMC. Although one may consider more advanced Metropolis-Hastings algorithms such as Hamiltonian Monte Carlo \citep{barbu2020hamiltonian}, they typically require an excessively long chain for burn-in and iterations till convergence when $p$ is large. The EM algorithm we develop here is similar to those implemented in \citet{rovckova2014emvs} and \citet{rovckova2018spike}; but instead of mean regression with Gaussian model error considered in these works, we adapt the algorithm to modal regression with MixHat model error. EM algorithms were initially motivated by maximum likelihood estimation in the presence of missing data. Linking this concept with our joint posterior distribution, we view $\boldsymbol{\lambda}$ as missing data and  \eqref{eq:JointPosterior} as the complete data likelihood with model parameters in $\Theta_{-\boldsymbol{\lambda}}=(\beta_0, \boldsymbol{\beta}^\top, \nu, \gamma, \theta)^\top$. We iterate between the E-step and M-step until the $L_2$-norm difference between the estimates for $\Theta_{-\boldsymbol{\lambda}}$ from the two successive iterations is less than $10^{-7}$.

{\bf\emph{The E-step}:} 
Starting with an initial estimate of $\Theta_{-\boldsymbol{\lambda}}$ denoted by $\Theta_{-\boldsymbol{\lambda}}^{(0)}$, at the E-step of the $(k+1)$-th iteration of the EM algorithm, for $k\ge 0$, we formulate the objective function defined as the expectation of the logarithm of \eqref{eq:JointPosterior} with respect to  $\boldsymbol{\lambda}$ given $\boldsymbol{y}$ and the current estimate of $\Theta_{-\boldsymbol{\lambda}}$, $\Theta^{(k)}_{-\boldsymbol{\lambda}}$. This leads to the following objective function, 
\begin{align}
&\ Q\left(\Theta_{\boldsymbol{-\lambda}} \left\vert \Theta^{(k)}_{-\boldsymbol{\lambda}}\right. \right) \notag \\
= &\ \sum_{i=1}^n 
\log \left\{ \left(1+\frac{\varepsilon_i^2}{\gamma^2 \nu} \right)^{-(\nu+1)/2} \mathbbm{1}_{[0,+\infty)}(\varepsilon_i)+ \left(1+\frac{\gamma^2\varepsilon_i^2}{\nu} \right)^{-(\nu+1)/2} \mathbbm{1}_{(-\infty,0)}(\varepsilon_i)\right\}-\frac{\beta_0^2}{2\times 10^6}+ \notag \\
&\ \sum_{j=1}^p \left( \log\frac{t_0}{2}- t_0\abs{\beta_j} \right)+ \sum_{j=1}^p E\left( \lambda_j \left \vert \beta_j^{(k)}, \theta^{(k)}, \boldsymbol{y}\right.\right) \left\{ \log\frac{t_1}{t_0}-  (t_1- t_0) \abs{\beta_j}\right\}+ \notag \\
&\ \log\left(\frac{\theta}{1-\theta}\right)\sum_{j=1}^p E\left(\lambda_j \left \vert \beta_j^{(k)}, \theta^{(k)}, \boldsymbol{y}\right.\right)  +  (a-1) \log\theta+ (b-1+p) \log(1-\theta)- \frac{(\log\nu - 1)^2}{2}  \notag \\
& - \left(\frac{n}{2}+1\right)\log\nu+ n\log\left(\frac{\Gamma((\nu+1)/2)}{\Gamma(\nu/2)}\right)+ (n+c-1) \log\gamma - n\log(\gamma^2+ 1)-d\gamma, 
\label{eq:Objective function}
\end{align}
where
\begin{equation}
\label{eq:Expectation}
E\left(\lambda_j \left\vert \beta_j^{(k)}, \theta^{(k)}, \boldsymbol{y}\right.\right) = \frac{1}{\displaystyle{1+ \frac{t_0}{t_1} \frac{(1-\theta^{(k)})}{\theta^{(k)}}\exp \left\{ -(t_0-t_1)\left\vert \beta_j^{(k)} \right\vert  \right\}} },
\end{equation}
which depends on $\Theta_{-\boldsymbol{\lambda}}^{(k)}$ via $\beta_j^{(k)}$  and $\theta^{(k)}$, as well as $t_0$ and $t_1$, and can be viewed as the posterior probability of including the $j$-th covariate conditional on $\beta_j^{(k)}$  and $\theta^{(k)}$ from the $k$-th iteration. 

{\bf \emph{The M-step}:}  
At the M-step of the $(k+1)$-th iteration of the EM algorithm, for $k\ge 0$, we update the estimate for $\Theta_{-\boldsymbol{\lambda}}$ by maximizing the objective function in \eqref{eq:Objective function} with respect to $\Theta_{-\boldsymbol{\lambda}}$, in the order $\boldsymbol{\beta}$, $\beta_0$, $\nu$, $\gamma$, and $\theta$. We first update the estimate for $\boldsymbol{\beta}$ by maximizing $Q(\Theta_{-\boldsymbol{\lambda}}|\Theta_{-\boldsymbol{\lambda}}^{(k)})$ in \eqref{eq:Objective function} with respect to $\boldsymbol{\beta}$ to obtain $\boldsymbol{\beta}^{(k+1)}$, which is equivalent to maximizing the following expression with respect to $\boldsymbol{\beta}$ while setting other parameters in $\Theta_{-\boldsymbol{\lambda}}$ at their most recent updates,
\begin{align}
&\ \sum_{i=1}^n 
\log \left\{ \left(1+\frac{\varepsilon_{i,k}^2}{\gamma^{(k)2} \nu^{(k)}} \right)^{-(\nu^{(k)}+1)/2} \mathbbm{1}_{[0,+\infty)}(\varepsilon_{i,k})+ \left(1+\frac{\gamma^{(k)2}\varepsilon_{i,k}^2}{\nu^{(k)}} \right)^{-(\nu^{(k)}+1)/2} \mathbbm{1}_{(-\infty,0)}(\varepsilon_{i,k})\right\} \notag\\
&\ -\sum_{j=1}^p \left\{\left(1-\hat p_j^{(k)}\right) t_0+\hat p_j^{(k)}t_1 \right\}|\beta_j|, \label{eq:Update_beta}
\end{align}
where $\varepsilon_{i,k}=y_i - \beta_0^{(k)} - \boldsymbol{x}_i ^\top\boldsymbol{\beta}$, and $\hat p_j^{(k)}$ represents $E(\lambda_j \vert \beta_j^{(k)}, \theta^{(k)},\boldsymbol{y})$ in \eqref{eq:Expectation}. The maximizer of \eqref{eq:Update_beta}, i.e., $\boldsymbol{\beta}^{(k+1)}$, is similar to a penalized maximum likelihood estimate, with the log-likelihood of the modal regression model given in the first summation in \eqref{eq:Update_beta} and the $L_1$-penalty given in the second term in \eqref{eq:Update_beta} that resembles a LASSO penalty. For given $t_0$ and $t_1$ with $t_1 \ll t_0$ such that the penalty is dominated by $(1-\hat p_j^{(k)})t_0$, the penalty is reduced when the probability of including the $j$-th covariate is higher. This intuitively sensible penalty dynamics bridges Bayesian variable selection utilizing the spike-and-slab prior for $\boldsymbol{\beta}$ and frequentist variable selection based on the LASSO penalized log-likelihood function \citep{tibshirani1996regression}. When $t_0=t_1$, the contribution of $\hat p_j^{(k)}$ to updating $\boldsymbol{\beta}$ vanishes, and $t_0$ is the penalty applied to all $p$ candidate covariates. This brings the optimizer of \eqref{eq:Update_beta} back to the traditional LASSO solution with the penalty fixed once chosen. Letting $t_0\ne t_1$ with $t_1 < t_0$ can thus produce a more versatile variable selection procedure than its LASSO counterpart. We elaborate on our choice of $(t_0, t_1)$, comparing with the strategies in competing methods in Section~\ref{sec:compete}.

Updates for  $\beta_0$, $\nu$, $\gamma$, and $\theta$, along with a coordinate ascent algorithm to update $\boldsymbol{\beta}$ that greatly improves the computational efficiency of the EM algorithm, are provided in the Web Appendix.

\section{Testing-driven variable selection}\label{sec:variable selection}
\subsection{Existing strategies} 
In the initial proposed SSL  \citep{rovckova2014emvs}, the posterior probability of including the $j$-th covariate, $\hat p_j^{(k)}$ in \eqref{eq:Expectation}, for $j=1, \ldots, p$, at convergence of the EM algorithm is used for variable selection with a pre-specified threshold such as 0.5. A potential drawback of this strategy is that it can lead to a high chance of selecting a negligible covariate because the probability in \eqref{eq:Expectation} increases exponentially with the difference between $t_0$ and $t_1$. Although \citet{rovckova2018spike} later developed a thresholding strategy directly applied to $\boldsymbol{\beta}$ for selecting important covariates, this strategy is not applicable in our regression model due to the flexible model error that leads to a more complex log-likelihood function. In fact, both aforementioned strategies tend to be sensitive to the choice of $(t_0, t_1)$. 

In what follows, we propose a new strategy to separate important covariates from unimportant ones following obtaining a Bayes estimate for $\Theta_{-\boldsymbol{\lambda}}$ from the EM algorithm. Denote the estimate by $\hat \Theta_{-\boldsymbol{\lambda}}=(\hat \beta_0, \hat {\boldsymbol{\beta}}^\top, \hat \nu, \hat \gamma, \hat \theta)^\top$, where $\hat {\boldsymbol{\beta}}=(\hat \beta_1, \ldots, \hat \beta_p)^\top$ are the penalized estimates for the covariate effects, and $(\hat \nu, \hat \gamma)$ shed light on the underlying distribution of the response.

\subsection{The test statistic}
The new strategy builds upon a test of the null hypothesis  $H_0^{(j)}:\, \beta_j=0$, for each $j\in \{1, \ldots, p\}$. The test statistic is inspired by the unimodality of the model error distribution in \eqref{eq:error distribution}, which suggests that the density function of $\varepsilon$ has a much steeper slope in a neighborhood of zero (excluding zero) compared to that at the tails further away from the mode (as zero) where the curvature tends to be mild. 

Denote by $p'(\varepsilon|\hat\nu, \hat \gamma)$ and $p''(\varepsilon|\hat\nu, \hat \gamma)$ the first and second derivatives of $p(\varepsilon|\hat\nu, \hat \gamma)$ with respect to $\varepsilon$, capturing the slope and curvature of the density curve at $\varepsilon$, respectively. The proposed test statistic to test $H_0^{(j)}$, for $j\in \{1, \ldots, p\}$, is based on the change in slope (CiS) of the estimated error density at two different residuals as follows, 
\begin{equation}
\text{CiS}_j=\frac{1}{n}\sum_{i=1}^n \frac{\left\vert \left\{p'(\hat \varepsilon_i|\hat \nu, \hat \gamma) \right\}^2 - \left\{p'(\hat \varepsilon_{i,-j}|\hat \nu, \hat \gamma) \right\}^2\right\vert}{|p''(\hat \varepsilon_{i,-j}|\hat \nu, \hat \gamma)|+ \delta},
\label{eq:CiS}
\end{equation}
where  $\delta$ is a small positive constant (e.g., $10^{-3}$ in our study ) to avoid a zero denominator, and,  for $i=1, \ldots, n$, $\hat \varepsilon_i=y_i-\hat \beta_0-\boldsymbol{x}_i^\top \hat {\boldsymbol{\beta}}$ and $\hat \varepsilon_{i,-j}=y_i-\hat \beta_0-\boldsymbol{x}_{i,-j}^\top \hat {\boldsymbol{\beta}}$, in which $\boldsymbol{x}_{i,-j}$ results from replacing the $j$-th entry of $\boldsymbol{x}_i$ with zero.  

Under $H_0^{(j)}$, $\hat \varepsilon_i$ and $\hat \varepsilon_{i,-j}$ are expected to be similar, leading to a small numerator in the summand of \eqref{eq:CiS}, and are likely to fall in a neighborhood of zero where the curvature tends to be less mild, resulting in a larger denominator in the summand. Conversely, when $\beta_j\ne 0$, $\hat \varepsilon_{i,-j}$ tends to shift away from zero compared to $\hat \varepsilon_i$, which can result in a drastic change in the slope of the density curve when moving from $\hat \varepsilon_i$ to $\hat \varepsilon_{i,-j}$, with less curvature at the latter as $\hat \varepsilon_{i,-j}$ shifting further away from zero. Consequently, a larger $\text{CiS}_j$ provides stronger data evidence against $H_0^{(j)}:\, \beta_j=0$. 

Because $\text{CiS}_j$ is defined through the relative change in slope when the $j$-th covariate is included versus excluded in the regression model, variable selection based on it tends to be much less sensitive to the choice of $(t_0, t_1)$ compared to the aforementioned existing strategies for variable selection. In particular, when the posterior distribution is primarily driven by the data likelihood, as typically seen in scenarios with $p < n$, we observed in our simulation study that the operating characteristics of the test statistic are fairly robust to variations in $(t_0, t_1)$. Certainly, when parameter estimation becomes more challenging and is subject to higher variability, as is often the case when $p > n$, some careful tuning for $(t_0, t_1)$ can help in parameter estimation, which in turn yields more reliable hypothesis testing based on $\text{CiS}_j$.

\subsection{A permutation test}
\label{subsec:permutation test}
To assess the statistical significance of $\text{CiS}_j$, we propose a procedure to estimate the $p$-value associated with an observed $\text{CiS}_j$ based on $B$ random permutations of the $j$-th covariate data. Following each permutation, we refit the modal regression model and compute the test statistic based on the permuted data. Such a permutation is expected to eliminate the potential association between the $j$-th covariate and the response, and thus create (permuted) data consistent with $H_0^{(j)}$. Viewing the $B$ realizations of the test statistic based on permuted datasets as references of what to expect under $H_0^{(j)}$, an estimated $p$-value is given by the relative frequency of these $B$ realizations exceeding the value of $\text{CiS}_j$ based on the original data. We then conclude inclusion of the $j$-th covariate only if the estimated $p$-value is less than a nominal level $\alpha$. We set $\alpha= 0.05$ in our study as most commonly done in frequentist hypothesis testing.  

Using the EM algorithm developed in Section~\ref{subsec:E-M algorithm}, the permutation test involving estimation of $\Theta_{-\boldsymbol{\lambda}}$ repeatedly is computationally manageable even when $p>n$. To further reduce the computational burden when $p$ is large, we add two pre-screening steps as follows. 
\begin{enumerate}[Step (i)]
    \item Randomly partition $p$ covariates into (nearly) equal-size groups, say, each of size $m$. Denote by $\boldsymbol{\beta}^{(q)}$ the regression coefficients associate with the $q$-th group of covariates, for $q=1, \ldots, \ceil{p/m}$. Test $H_0^{(q)}:\, \boldsymbol{\beta}^{(q)}=\boldsymbol{0}$ based on a test statistic similar to \eqref{eq:CiS} defined by 
    $$
\text{CiS}_{(q)}=\frac{1}{n}\sum_{i=1}^n \frac{\left\vert \left\{p'(\hat \varepsilon_i|\hat \nu, \hat \gamma) \right\}^2 - \left\{p'(\hat \varepsilon_{i,-(q)}|\hat \nu, \hat \gamma) \right\}^2\right\vert}{|p''(\hat \varepsilon_{i, -(q)}|\hat \nu, \hat \gamma)|+ \delta},
$$ 
where $\hat\varepsilon_{i, -(q)}=y_i-\hat \beta_0- \boldsymbol{x}_{i,-(q)}^\top \hat {\boldsymbol{\beta}}$, in which $\boldsymbol{x}_{i,-(q)}$ results from replacing the covariates values corresponding the group-$q$ covariates with zeros. A  permutation procedure with a small number of permutations, say, $B_1=20$, is used to assess the significance of $\text{CiS}_{(q)}$ for each $q \in \{1, \ldots, \ceil{p/m}\}$. In each permutation, the rows of the $n\times m$ submatrix of $\boldsymbol{X}$ corresponding to the considered group of covariates are permuted. The $q$-th group of covariates is concluded as insignificant if the estimated $p$-value associated with $\text{CiS}_{(q)}$ falls above $\alpha_0$. We set $\alpha_0=6/20$ in our study. 
    \item Suppose $p_1$ candidate covariates remain in the regression model after concluding $n-p_1$ covariates insignificant at Step (i). Carry out the permutation procedure described above for testing the significance of one covariate at a time among the remaining $p_1$ covariates based on \eqref{eq:CiS} with a small number of permutations, say, $B_2=20$. Conclude a covariate is insignificant if the resultant $p$-value is above $\alpha_0$.   
\end{enumerate}
Suppose $p_2$ candidate covariates remain in the model after Step (ii). We then repeat the permutation procedure based on the test statistic in \eqref{eq:CiS} for this (usually much smaller) collection of $p_2$ candidate covariates using a large number of permutations, say, $B=200$, with the threshold on $p$-value returning to $\alpha(=0.05)$.

Step (i) carries out group-level screening that aims to remove an entire group of covariates from the model if there exists no influential covariate in the group. Although the group size can impact the efficiency in screening at this step, we find in our empirical study with a large $p$ that a group size of $m=4$ often provides a good balance between computing time and screening efficiency. Step (ii) performs individual-level screening where we further screen out unimportant covariates one at a time based on a small number of permutations for each of the $p_1$ remaining covariates from Step (i). The relatively small number of permutations (e.g., $B_1=B_2=20$) and the nominal significance level (e.g., $\alpha_0 = 6/20$) we use in Steps (i) and (ii) are inspired by the pretesting procedure described in \citet{davidson2000bootstrap}. Following our design of the two-step screening, the probability of concluding a truly non-negligible covariate insignificant is merely around $3 \times 10^{-5}$ when assuming a false discovery rate of 0.05. 

In the sequel, we refer to the variable selection procedure involving the proposed permutation test following Bayesian modal regression analysis as the {\underline t}esting-{\underline d}riven \underline{v}ariable \underline{s}election, or, TDVS for short.

\section{Simulation studies}\label{sec:Simulations}
We now inspect the finite sample performance of TDVS in designed simulation experiments where response data can be subject to severe outliers.  
\subsection{Design of simulation experiments}\label{subsec:simulation scenarios}
There are three varying factors in the experiments. One factor is the value of $(p, n) \in \{(8, 100), (80, 30)\}$, providing a contrast between a small $p$ and a large $p$ relative to $n$. The other factor is the model error distribution, for which we consider three distributions: (i) $\text{MixHat}(\nu=3, \gamma=2)$, which is right-skewed with an excess kurtosis of infinity, (ii) $\mathcal{N}(0, 3)$, which can be well approximated by $\text{MixHat}(\nu, 1)$ with a large enough $\nu$, and (iii) the Gaussian mixture  $0.8\mathcal{N}(0, 3)+0.2 \mathcal{N}(5, 7)$ as a mode-zero right-skewed distribution with the excess kurtosis equal to 0.744. The third factor is the distribution of covariates. We assume that they are normally distributed, each with mean $0$, and allow certain covariates to be correlated. Common in all experiments are the intercept $\beta_0=2$ and the only two non-null covariate effects, $\beta_1=2$ and $\beta_3=1$, in the regression model \eqref{eq:Linear Regression}. Under each simulation setting, we simulate 300 Monte Carlo replicate data sets with the response data generated based on \eqref{eq:Linear Regression}.  

Four metrics are used to assess the performance of a considered method: (i) the true positive rate (TPR) as the proportion of all truly non-negligible covariate effects being identified, (ii) the false positive rate (FPR) as the proportion of all truly null covariate effects being concluded as non-negligible, (iii) the accuracy rate (ACR) as the proportion of correct identification, either as non-negligible or negligible covariate effects, across $p$ candidate covariates, and (iv) the mean squared error (MSE) of $\hat{\boldsymbol{\beta}}$, defined as $\|\hat{\boldsymbol{\beta}}- \boldsymbol{\beta}\|^2_2/p$. 

\subsection{Competing methods}
\label{sec:compete}
As close relatives to our proposed method, we include the traditional LASSO and SSL, two mean regression models, as competing methods. To implement the former, we use the \texttt{R} package \texttt{glmnet} \citep{glmnet2010}, which carries out cross-validation based on prediction mean squared error (PMSE) to choose $t_0$. To implement SSL, we use the \texttt{R} package \texttt{SSLASSO} that executes the dynamic posterior exploration, where one updates the estimate for $\Theta$ iteratively over an ascending sequence of $t_0$ with a fixed $t_1$, until the maximum $t_0$ (default $n$) is reached. In essence, this strategy does not attempt to choose an optimal $t_0$ but instead sequentially increases $t_0$, during which the estimation of $\Theta$ stabilizes.

For TDVS, the choice of $(t_0, t_1)$ is much less impactful on the selection outcomes than for the competing methods when $p<n$, partly thanks to the test based on CiS. For instance, as reported in the simulation results (Subsection~\ref{subsec:Simulation Results}), when $p$ is small relative to $n$, varying $t_0$ while fixing $t_1=1$ does not alter inference results noticeably. When $p > n$, we set $t_1=1$ and choose $t_0$ via a five-fold cross-validation based on PMSE to ensure estimation stability and encourage sparse models.

\subsection{Simulation results}\label{subsec:Simulation Results}
Table~\ref{tab:variable selection results with small p} presents summary statistics for each of the considered methods when $(p, n)=(8, 100)$ and the covariates data are generated from $\mathcal{N}(\boldsymbol{0}_8, \boldsymbol{I}_8)$. To demonstrate the robustness of TDVS to the choice of $t_0$, we varied $t_0$ over a wide range of values while fixing $t_1$ at 1. The robustness in variable selection is persistent across different model error distributions. This showcases the benefit of the proposed testing procedure that makes TDVS a nearly tuning-free method when $p$ is not large. LASSO and SSL slightly outperform TDVS overall when the model error is Gaussian, although TDVS achieves a higher TPR than SSL. This is expected since they are designed for mean regression models with Gaussian errors. TDVS substantially outshines LASSO and SSL in the presence of non-Gaussian model error, with a much higher TPR, well-controlled FPR, and an improved ACR in variable selection, as well as significantly lower MSE in covariate effects estimation. This is the case even when TDVS misspecifies the model error, that is, assuming a (unimodal) $\text{MixHat}$ distribution when the truth is a (nearly bimodal) Gaussian mixture. The FPR of TDVS is close to 0.05, as a result of our design of the permutation test at a significance level of 0.05, which in turn indicates that the proposed test preserves the nominal test size very well. In contrast, SSL tends to yield a lower FPR, suggesting its conservative nature in variable inclusion. 

\begin{table}[htbp]
\caption{Averages of TPR, FPR, ACR, and MSE across 300 Monte Carlo replicates for three considered methods when $p=8$ and $n=100$, with covariates data generated from $\mathcal{N}(\boldsymbol{0}_p, \boldsymbol{I}_p)$. Numbers in parentheses are Monte Carlo standard errors associated with the averages.}
\label{tab:variable selection results with small p}
\centering
\begin{tabular}{lcccc}
\hline 
Method & TPR & FPR & ACR & MSE  \\
\hline
& \multicolumn{4}{c}{$\varepsilon\sim \text{MixHat}(3, 2)$} \\
\cline{2-5}
TDVS ($t_0=1$) & 0.990 (0.004) & 0.037 (0.005) & {\bf 0.970} (0.004) & 0.020 (0.001) \\
TDVS ($t_0=10$) & {\bf 1.000} (0.000) & 0.048 (0.005) & 0.964 (0.004) & {\bf 0.007} (0.001)  \\
TDVS ($t_0=30$) & 0.993 (0.003) & 0.046 (0.005) & 0.964 (0.004) & 0.009 (0.001)  \\
TDVS ($t_0=100$) & 0.972 (0.007) & 0.042 (0.005) & 0.962 (0.004) & 0.023 (0.003)  \\
LASSO & 0.723 (0.021)  & 0.005 (0.002)  & 0.927 (0.005) & 0.249 (0.011)  \\
SSL & 0.798 (0.015)  & {\bf 0.001} (0.001)  & 0.949 (0.004) & 0.068 (0.006)  \\
\cline{2-5}
& \multicolumn{4}{c}{$\varepsilon\sim \mathcal{N}(0, 3)$} \\ 
\cline{2-5}
TDVS ($t_0=1$) & 0.952 (0.009) & 0.032 (0.005) & 0.964 (0.004) & 0.040 (0.001)  \\
TDVS ($t_0=10$) & 0.990 (0.004) & 0.039 (0.005)  & 0.968 (0.004) & {\bf 0.018} (0.001)  \\
TDVS ($t_0=30$) & 0.985 (0.005) & 0.044 (0.005) & 0.963 (0.004) & 0.019 (0.001)  \\
TDVS ($t_0=100$) & 0.962 (0.008) & 0.039 (0.005) & 0.961 (0.004) & 0.031 (0.002)  \\
LASSO & {\bf 0.995} (0.003)  & 0.024 (0.004)  & 0.980 (0.003) & 0.065 (0.002)  \\
SSL & 0.930 (0.010)  & {\bf 0.000} (0.000) & {\bf 0.983} (0.003) & 0.024 (0.002)  \\
\cline{2-5}
& \multicolumn{4}{c}{$\varepsilon\sim 0.8 \mathcal{N}(0, 3)+0.2\mathcal{N}(5, 7)$} \\
\cline{2-5}
TDVS ($t_0=1$) & 0.838 (0.014)  & 0.048 (0.005)  & 0.923 (0.005) & 0.074 (0.003)  \\
TDVS ($t_0=10$) & {\bf 0.970} (0.007)  & 0.046 (0.005)  & {\bf 0.958} (0.004) & {\bf 0.035} (0.002)  \\
TDVS ($t_0=30$)& 0.923 (0.010) & 0.045 (0.005) & 0.947 (0.005) & 0.040 (0.003)  \\
TDVS ($t_0=100$)& 0.890 (0.012) & 0.042 (0.005) & 0.941 (0.005) & 0.053 (0.003)  \\
LASSO & 0.815 (0.014)  & 0.011 (0.002)  & 0.946 (0.004) & 0.197 (0.006)  \\
SSL & 0.723 (0.014)  & {\bf 0.001} (0.001)  & 0.930 (0.004) & 0.082 (0.004)  \\
\hline
\end{tabular}
\end{table}

Table~\ref{tab:variable selection results with correlated predictors} provides the counterpart results from the same simulation settings as for Table~\ref{tab:variable selection results with small p} except that the covariate data are generated from $\mathcal{N}(\boldsymbol{0_8}, \boldsymbol{\Sigma})$, where $\boldsymbol{\Sigma}=\text{diag}(\boldsymbol{\Sigma}_1,\boldsymbol{\Sigma}_2,\boldsymbol{\Sigma}_3, \boldsymbol{\Sigma}_4)$, in which $\boldsymbol{\Sigma}_k=0.5\boldsymbol{J}_2+0.5\boldsymbol{I}_2$, $k \in \{1,2,3,4\}$, with $\boldsymbol{J}_2$ denoting an $2 \times 2$ matrix of one's. This design of correlated covariates creates a challenging setting for most well-accepted variable selection methods due to violating the irrepresentable conditions \citep{zhao2006model}, under which an important covariate can only weakly correlate with unimportant covariates compared to its correlation with important covariates. Regardless, TDVS maintained a better balance between avoiding false selection and identifying truly important covariates than the competing methods when the model errors are non-Gaussian, although the choice of $t_0$ can influence the performance more noticeably. The correlation between covariates also contributes to the higher (than 0.05) FPR for TDVS because, by permuting data for one covariate at a time in the permutation test, the permuted data distort the original correlation structure, making it harder to accurately recover the null distribution of CiS based on the permuted data.

\begin{table}[htbp]
\centering
\caption{Averages of TPR, FPR, ACR, and MSE across 300 Monte Carlo replicates for three considered methods when $p=8$ and $n=100$, with covariates data generated from $\mathcal{N}(\boldsymbol{0}_p, \boldsymbol{\Sigma})$, where $\boldsymbol{\Sigma}\ne \boldsymbol{I}_p$. Numbers in parentheses are Monte Carlo standard errors associated with the averages.}
\label{tab:variable selection results with correlated predictors}
\centering
\begin{tabular}{lcccc}
\hline 
Method & TPR & FPR & ACR & MSE \\
\hline
& \multicolumn{4}{c}{$\varepsilon\sim \text{MixHat}(3, 2)$} \\
\cline{2-5}
TDVS ($t_0=1$) & 0.930 (0.011) & 0.063 (0.006) & 0.935 (0.005)  & 0.024 (0.001)  \\
TDVS ($t_0=10$) & {\bf 0.998} (0.002) & 0.044 (0.005) & {\bf 0.967} (0.004)  & {\bf 0.009} (0.001)  \\
TDVS ($t_0=30$) & 0.940 (0.010) & 0.072 (0.007) & 0.931 (0.006)   & 0.036 (0.003)  \\
TDVS ($t_0=100$) & 0.818 (0.015) & 0.063 (0.007) & 0.907 (0.006)   & 0.162 (0.009)  \\
LASSO & 0.710 (0.021)  & 0.012 (0.003)  & 0.919 (0.005) & 0.251 (0.011)  \\
SSL & 0.782 (0.015)  & {\bf 0.001} (0.001)  & 0.945 (0.004)  & 0.071 (0.005)  \\
\cline{2-5}
& \multicolumn{4}{c}{$\varepsilon\sim \mathcal{N}(0, 3)$} \\ 
\cline{2-5}
TDVS ($t_0=1$) & 0.895 (0.014) & 0.060 (0.007) & 0.929 (0.006)   & 0.052 (0.002)  \\
TDVS ($t_0=10$) & 0.987 (0.005) & 0.057 (0.006)  & 0.954 (0.004)  & 0.025 (0.001)   \\
TDVS ($t_0=30$) & 0.952 (0.009) & 0.063 (0.006) & 0.940 (0.005)   & 0.059 (0.002)  \\
TDVS ($t_0=100$) & 0.942 (0.010) & 0.064 (0.007) & 0.938 (0.005)   & 0.077 (0.003)  \\
LASSO & {\bf 0.998} (0.002)  & 0.053 (0.006)  & 0.960 (0.005) & 0.065 (0.002)  \\
SSL & 0.942 (0.009)  & {\bf 0.001} (0.001) & {\bf 0.985} (0.002) & {\bf 0.021} (0.002)  \\
\cline{2-5}
& \multicolumn{4}{c}{$\varepsilon\sim 0.8 \mathcal{N}(0, 3)+0.2\mathcal{N}(5, 7)$} \\
\cline{2-5}
TDVS ($t_0=1$) & 0.742 (0.017) & 0.051 (0.006) & 0.897 (0.006)  & 0.091 (0.003)  \\
TDVS ($t_0=10$) & {\bf 0.890} (0.013)  & 0.067 (0.007)  & 0.922 (0.006)  &  {\bf 0.056} (0.003)  \\
TDVS ($t_0=30$) & 0.787 (0.016) & 0.062 (0.006) & 0.900 (0.006)  & 0.107 (0.005)  \\
TDVS ($t_0=100$) & 0.777 (0.016) & 0.065 (0.007) & 0.895 (0.006)  & 0.124 (0.005)  \\
LASSO & 0.797 (0.015)  & 0.021 (0.004)  & {\bf 0.933} (0.004)  & 0.205 (0.006)  \\
SSL & 0.708 (0.014)  & {\bf 0.001} (0.001)  & 0.927 (0.004)  & 0.089 (0.004)  \\
\hline
\end{tabular}
\end{table}

Lastly, Table~\ref{tab:variable selection results with large p} shows results under the setting with $(p, n)=(80, 30)$. With $p>n$ here, we tuned $t_0$ and implemented the pre-screening steps described in Section~\ref{subsec:permutation test}. The TPRs of all three methods in this large-$p$-small-$n$ setting are much lower than those in Tables~\ref{tab:variable selection results with small p} and \ref{tab:variable selection results with correlated predictors}. Still, TDVS achieves the highest TPR among the three while remaining an FPR close to our pre-set nominal level of 0.05 when the model error substantially deviates from Gaussian, the type of data that modal regression is especially suitable for. 

Based on the simulation findings, in all real data applications, we set $(t_0, t_1)=(10,1)$ as the default choice when $p < n$, and tune $t_0$ with a fixed $t_1=1$ via five-fold cross-validation based on PMSE when $p > n$.

\begin{table}[htbp]
\caption{Averages of TPR, FPR, ACR, and MSE across 300 Monte Carlo replicates for three considered methods when $p=80$ and $n=30$, with covariates data generated from $\mathcal{N}(\boldsymbol{0}_p, \boldsymbol{I}_p)$. Numbers in parentheses are Monte Carlo standard errors associated with the averages.}.
\label{tab:variable selection results with large p}
\centering
\begin{tabular}{lccccc}
\hline
Method & TPR & FPR & ACR & MSE \\
\hline
& \multicolumn{4}{c}{$\varepsilon\sim \text{MixHat}(3, 2)$} \\
\cline{2-5}
TDVS & {\bf 0.617} (0.021) & 0.038 (0.002) & 0.953 (0.002) &  {\bf 0.031} (0.002) \\
LASSO & 0.460 (0.023)  & 0.016 (0.002)  & 0.971 (0.002) & 0.041 (0.001) \\
SSL & 0.308 (0.014)  & {\bf 0.000} (0.000)  & {\bf 0.983} (0.000) & 0.034 (0.001) \\
\cline{2-5}
& \multicolumn{4}{c}{$\varepsilon\sim \mathcal{N}(0, 3)$} \\
\cline{2-5}
TDVS & 0.620 (0.021) & 0.037 (0.002) & 0.955 (0.002) & 0.030 (0.001)  \\
LASSO & {\bf 0.695} (0.018)  & 0.027 (0.003)  & 0.966 (0.003) & 0.028 (0.001) \\
SSL & 0.427 (0.010)  & {\bf 0.000} (0.000) & {\bf 0.986} (0.000) &  {\bf 0.022} (0.001)  \\
\cline{2-5}
& \multicolumn{4}{c}{$\varepsilon\sim 0.8 \mathcal{N}(0, 3)+0.2\mathcal{N}(5, 7)$} \\
\cline{2-5}
TDVS & {\bf 0.378} (0.019) & 0.036 (0.002) & 0.950 (0.002) & 0.065 (0.003)  \\
LASSO & 0.285 (0.020)  & 0.015 (0.002)  & 0.968 (0.002) &  0.051 (0.001)  \\
SSL & 0.157 (0.013)  & {\bf 0.000} (0.000)  & {\bf 0.979} (0.000) & {\bf 0.048} (0.001)  \\
\hline
\end{tabular}
\end{table}

\section{Applications in medical research}\label{sec:Real data examples}
\subsection{Genetics in acute myeloid leukemia}
Acute myeloid leukemia is a cancer of the blood and bone marrow. Genes associate with the disease through genetic mutations  \citep{lagunas2017acute}. Examples of these genes include \textit{FLT3}, \textit{CEBPA}, \textit{NPM1}, and \textit{KIT}, with the former three also found to relate to treatment response \citep{takahashi2011current}. Lymphocyte-activation gene 3 (\textit{LAG3}) is a marker for T cell exhaustion and tumor progression, which can promote myeloid-leukemia-induced immune suppression \citep{abdelhakim2018lag3}. For illustration purposes, we view the expression level of \textit{FLT3} as the response and study its potential association with \textit{CEBPA}, \textit{NPM1}, \textit{KIT}, and \textit{LAG3} based on gene expression data of $n=151$ patients diagnosed with acute myeloid leukemia from the Genomic Data Commons (\url{https://portal.gdc.cancer.gov/}). 
In addition to gene expression measurements, the data also include patients' demographic information, the following of which we include as covariates in our analysis: gender, race, and age. 
This gives a total of $p=7$ candidate covariates. 

\begin{figure}[htbp]
    \centering
    \subfigure[]{
    \includegraphics[width=0.45\textwidth]{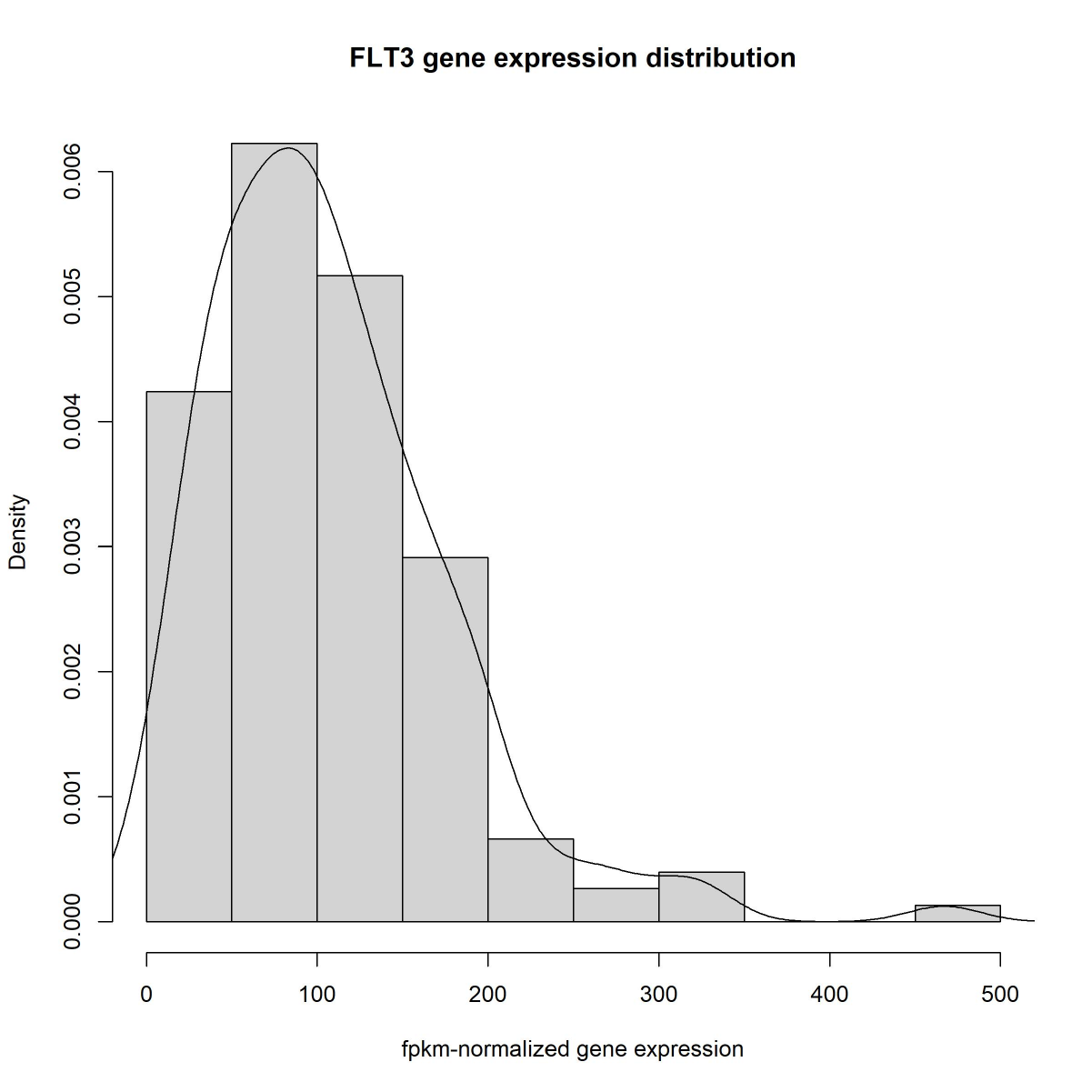}
        \label{fig:real data application1}
    }
    \hfill
    \subfigure[]{
        \includegraphics[width=0.45\textwidth]{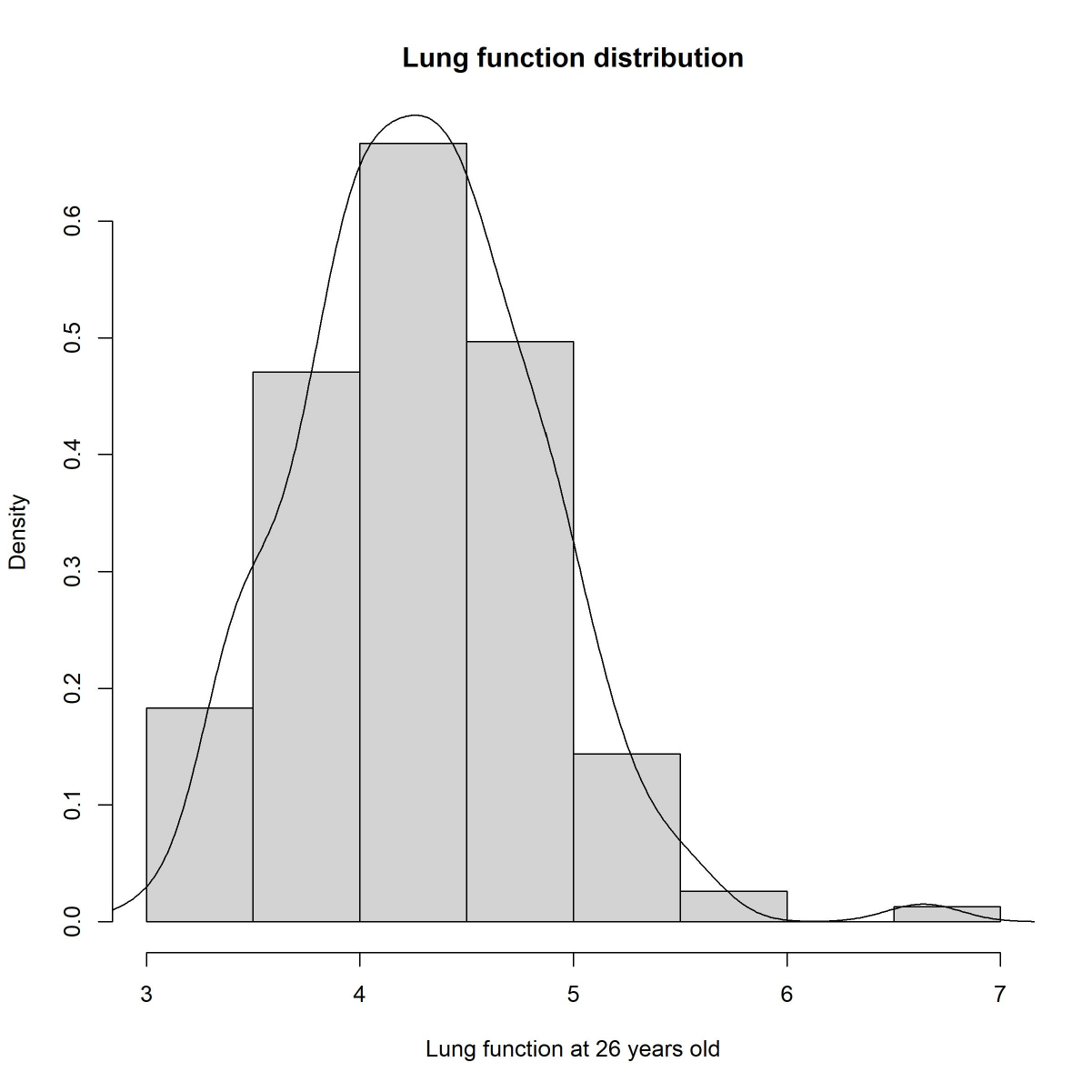}
        \label{fig:real data application2}
    }
    \caption{Histograms of response data analyzed in Section~\ref{sec:Real data examples}, superimposed with kernel density estimates}
    \label{fig:Response distributions in real data applications}
\end{figure}

Figure~\ref{fig:real data application1} shows the histogram of the observed \textit{FLT3} expression levels, which reveals the existence of severe outliers suggesting an underlying right-skewed distribution for the response that is potentially heavy-tailed. This observation motivates our modal regression model for the \textit{FLT3} expression level formulated in Section~\ref{sec:The regression model}. Applying TDVS to the data with $(t_0, t_1)=(10, 1)$, we concluded that \textit{CEBPA}, \textit{NPM1}, and \textit{KIT} as influential covariates for \textit{FLT3}. The estimated error distribution is $\text{MixHat}(\hat \nu=0.228, \hat \gamma=0.886)$, which aligns with the heavy tail of the empirical observation in Figure~\ref{fig:real data application1}. The $\gamma$ estimate indicates a left-skewed model error distribution, while the response distribution exhibits right skewness, a phenomenon that is plausible in regression models. In contrast, under the assumption of a Gaussian model error, LASSO and SSL selected a null model with none of the candidate covariates included as influential factors for the response. These results are reminiscent of the observed pattern in the simulation study that TDVS tends to possess a stronger potential to detect important covariates than LASSO and SSL in the presence of heavy-tailed response data.

\subsection{Epigenetic study for lung function}
\label{subsec:lung function}
We now turn to another example that examines the association between DNA methylation and lung function measured by forced vital capacity (FVC). The data are from the Isle of Wight birth cohort in the United Kingdom \citep{arshad2018cohort}, where DNA methylation in whole blood at genome-scale cytosine-guanine (CpG) sites was assessed at different ages of a cohort of participants. In our analyses, CpG sites with DNA methylation at age 18 years are included as covariates to establish an association with an individual's FVC measured at 26 years of age.  DNA methylation has been quality controlled, pre-processed, and cell type-adjusted  \citep{campbell2020cell}.
Figure~\ref{fig:real data application2} depicts the histogram of the FVC of $n=153$ participants at age 26 years, which appears to be slightly right-skewed with a few outliers much less severe than those in the first example shown in Figure~\ref{fig:real data application1}. 

Considering the tremendously large number of CpG sites, we started with a reduced set of 122 CpG sites identified to be informative by an \texttt{R} package, \texttt{ttScreening} \citep{ray2016efficient}. Following \texttt{ttScreening}, one quantifies the potential connection of each CpG site with lung function via robust linear regression, producing a connection score for each CpG site that increases with the strength of its association with the response. Based on this reduced set of CpG sites, we conducted three rounds of analyses to compare  TDVS, LASSO, and SSL under different settings.

In the first analysis, we picked four CpG sites with the highest connection scores, cg24216596, cg08770757, cg08244750, cg21938179, and randomly selected four other CpG sites, cg04941246, cg00000109, cg00000165, cg27170782, deemed uninformative according to \texttt{ttScreening} as candidate predictors. With $(t_0, t_1)=(10, 1)$, TDVS selected a model with five covariates, including DNAm of all top four CpG sites according to their connection scores, and one other site (cg04941246). LASSO selected the same model. As a byproduct of the analysis from TDVS, the estimated model error distribution is $\text{MixHat}(\hat\gamma=1.016, \, \hat \nu=17.225)$, as a nearly symmetric and light-tailed distribution. In contrast, SSL chose only two of the top four CpG sites (cg08770757, cg08244750) and one other site (cg04941246) to be included as influential covariates. The selection of CpG site cg04941246 by all three methods is likely due to its moderate correlation with the CpG site cg24216596 that has the highest connection score, with Pearson correlation coefficient of 0.452. In this application, the finding is in line with our observation in the simulation study in that, when the error distribution is (nearly) Gaussian, TDVS and LASSO tend to agree, while SSL is more conservative in claiming influential covariates.  

In the second round of analyses, we randomly permuted the top four CpG sites' data across the 153 subjects, and use the permuted data as DNAm values for four hypothetical CpG sites. These four hypothetical sites and the top four CpG sites constitute the set of candidate covariates in this analysis, where we expect the four hypothetical ones to be unimportant covariates for the response. Based on this new dataset, TDVS (with $t_0= 10$ and $t_1=1$) and LASSO again selected the same model that only includes DNAm of the top four CpG sites as important covariates. Additionally, the estimated error distribution according to TDVS is $\text{MixHat}(\hat \gamma=1.017, \, \hat \nu=17.295)$, similar to the one based on the raw data in the first round of analysis. As a conservative model selector, SSL included three of the top four CpG sites as influential covariates.

Lastly, we created a scenario with $p > n$ in the third round of analysis. In particular, we randomly picked 61 CpGs from the 122 CpGs that are informative according to \texttt{ttScreening}; we then permuted the data associated with these 61 CpGs to generate data for 100 pseudo CpGs. This produces a dataset with $p=161$ candidate covariates and $n = 153$ observations for subsequent variable selection. After the pre-screening steps described in Section~\ref{subsec:permutation test} and then tuning $t_0$ (to be 1) using cross-validation based on PMSE, TDVS selected 38 of the original 61 CpG sites along with 16 of the 100 pseudo CpGs. Moreover, the estimated error distribution according to TDVS is $\text{MixHat}(\hat \gamma=0.996, \, \hat \nu=15.212)$, which again suggests a nearly Gaussian error distribution. In comparison, LASSO selected 42 from the original 61 CpGs and 1 from the 100 pseudo CpGs, and SSL chose a model with only 2 CpG sites, all belonging to the original set of 61 sites. This resembles the trend observed in the simulation study when $p>n$ and the response data are (nearly) Gaussian: LASSO tends to discover more truly important covariates than TDVS while keeping the false discovery rate low.

\section{Discussion}\label{sec:disc}
We present in this study a flexible framework for variable selection that can accommodate
 various types of unimodal distributions for the response, including Gaussian and a wide range of distributions deviating from normality in symmetry or in the lightness of tails. The contributions of the proposed methodology are at least twofold. Firstly, we incorporate a flexible error distribution into Bayesian variable selection exploiting a spike-and-slab prior for covariates effects. Using the mode of the response as the central tendency measure given covariates allows for efficient prediction of the response with practically meaningful interpretations. The spike-and-slab prior yields a dynamic penalty in the posterior of model parameters used to conduct variable selection more effectively than the traditional LASSO. The EM algorithm, along with our proposed pre-screening procedure, accelerates the selection process and makes it more scalable than the conventional MCMC. Secondly, we propose the change-in-slope statistic to assess the significance of an inferred covariate effect, equipped with a permutation test. The testing-driven strategy for variable selection improves over SSL in regard to robustness of the selection outcomes to hyperparameters, especially when $p<n$, leading to a well-balanced TPR and FPR in identifying influential covariates. 

Variable selection with correlated candidate covariates, such that the irrepresentable conditions \citep{zhao2006model} are violated, presents challenges for all considered methods. But TDVS still performs competitively in such challenging scenarios. Existing methods designed to cope with correlated covariates are mostly developed for mean regression models from the perspective of partial correlation, where an individual covariate effect is assessed after controlling for effects of other covariates \citep{li2017variable, xue2022semi}. Similar techniques are not directly applicable for modal regression where covariate effects appear in the mode instead of the mean of the response, and certain moments may not exist for the potentially heavy-tailed model error to guarantee a well-defined partial correlation. Consistent variable selection based on modal regression when the irrepresentable conditions are violated remains an open problem for future research.

When $p$ is large enough to exceed $n$, the proposed method can benefit from a more carefully chosen $t_0$. We currently use a cross-validation based on PMSE to choose $t_0$, which we believe has room for improvement because PMSE is more suitable as a model criterion for mean regression. An interesting direction to pursue next is to use a different model criterion tailored for modal regression in cross-validation to tune $t_0$.   

Permutation-based tests are inherently time-consuming. We incorporated several improvements in the numerical implementation of TDVS, including pre-screening and the coordinate ascent algorithm, to reduce computational cost. One may consider different variants of permutation tests that are more computationally efficient, such as that in  \citet{leem2018approximation}, to estimate the $p$-value of the test statistic, CiS. Alternatively, one may propose a parametric distribution family indexed by a few parameters, such as a generalized gamma or generalized Pareto distribution, to approximate the null distribution of CiS so that only a small-scale permutation suffices to estimate the few parameters. 

An extension of the proposed variable selection method can be used in neighborhood selection for network analysis \citep{meinshausen2006high}, where a $p-\text{node}$ network can be specified by structural equation modeling that involves $p$ regression models. Given the limited advancements in network analysis for non-Gaussian data \citep{shojaie2021differential}, our approach can be a useful stepping stone to network analysis for heavy-tailed data.

\section*{Acknowledgments}

The study is partially supported by the National Institutes of Health, USA, under award numbers R21 AI175891 (HZ) and R01 AI121226 (HZ). The authors are grateful to Dr. Hasan Arshad at the University of Southampton for his support to this study and to the utilization of the DNA methylation and lung function data in this study.

\section*{Data availability statement}
\texttt{R} scripts for simulation data generation and analysis are available at \url{https://github.com/Jiasong-Duan/TDVS-simulations}. Acute Myeloid Leukemia data are included in the online supplementary materials and were extracted from the Genomic Data Commons at \url{https://portal.gdc.cancer.gov/}. DNA methylation and lung function data are available from the corresponding author upon reasonable request, with permissions from Dr. Hongmei Zhang and Dr. Hasan Arshad, respectively. 

\section*{Supplementary material}

The Web Appendix referenced in Sections~\ref{sec:Parameter estimation} is available online on the arXiv website.

%
%
\bibliographystyle{apalike}
\bibliography{Reference}

\end{document}